\documentclass[12pt,letterpaper,onecolumn,journal,draftcls]{IEEEtran}
\usepackage[latin9]{inputenc}
\usepackage{amsmath}
\usepackage{amssymb}
\usepackage{graphicx}
\usepackage{esint}

\makeatletter

\usepackage{amsfonts}
\usepackage{cite}
\usepackage{color}
\usepackage{epstopdf}

\providecommand{\U}[1]{\protect\rule{.1in}{.1in}}
\pdfoutput=1
\providecommand{\U}[1]{\protect\rule{.1in}{.1in}}

\newtheorem{theorem}{Theorem}
\newtheorem{lemma}{Lemma}

\newcommand{\qed}{\nobreak \ifvmode \relax \else
      \ifdim\lastskip<1.5em \hskip-\lastskip
      \hskip1.5em plus0em minus0.5em \fi \nobreak
      \vrule height0.75em width0.5em depth0.25em\fi}

\makeatother

\begin{document}

\title{Strictly Positive and Continuous Random Fibonacci Sequences and Network
Theory Applications}
\author{\IEEEauthorblockN{David Simmons and Justin P. Coon} \thanks{\IEEEauthorblockA{The authors are with the Department
of Engineering Science, University of Oxford, Parks Road, Oxford, OX1 3PJ, UK. 
 Email: $\{$david.simmons, justin.coon$\}$@eng.ox.ac.uk}}}
\maketitle
\begin{abstract}
We motivate the study of a certain class of random Fibonacci
sequences - which we call continuous random Fibonacci sequences - by demonstrating that their exponential growth rate can be used to establish capacity and power scaling laws for multihop cooperative amplify-and-forward (AF) relay networks. With these laws, we show that it is possible to construct multihop cooperative AF networks that simultaneously avoid 1) exponential capacity decay and 2) exponential transmit power growth across the network. This is achieved by ensuring  the network's Lyapunov exponent is zero. 
\end{abstract}

\section{Introduction}

With $F_{1}=F_{2}=1$,  the $n$th Fibonacci number
is defined to be $F_{n}=F_{n-1}+F_{n-2}$,  and is named after the Italian
mathematician Leonardo Fibonacci \cite{fibonacci2003fibonacci}. These
numbers frequently arise in nature \cite{douady1996phyllotaxis};
e.g., the leaf arrangement on a stem, the spiral number in
the floret of a sunflower, and the ancestry code of a male bee \cite{basin1963fibonacci}.
Fibonacci numbers are also connected to the golden ratio $\phi:=\left(1+\sqrt{5}\right)/2$,
which has been used to proportion the work of artists and architects
throughout history \cite{livio2008golden}. In particular, it is well
known that $\lim_{n\to\infty}\left(F_{n}/F_{n-1}\right)=\phi.$

\emph{Random} Fibonacci sequences were first introduced in 2000 \cite{viswanath2000random},
where it was shown that, with $x_{n}$ defined by 
\[
x_{n+1}=\pm x_{n}\pm x_{n-1}
\]
where the signs are chosen independently and with equal probability,
the limit 
\[
\lim_{n\to\infty}\left|x_{n}\right|^{\frac{1}{n}}\overset{a.s.}{=}C
\]
exists almost surely with $C=1.13198824...$ (Viswanath's constant).
The random Fibonacci sequence was generalized to $x_{n+1}=\pm x_{n}\pm\beta x_{n-1}$
$\beta>0$, and studied further in \cite{embree1999growth}, where
it was shown that there was a critical threshold $\beta^{\star}$
such that $\beta>\beta^{\star}$ would cause exponential growth in
the sequence, while $\beta<\beta^{\star}$ would cause exponential
decay. Further work was performed in \cite{janvresse2008random},
where the exponential growth rate of $x_{n+1}=x_{n}\pm x_{n-1}$
was assessed under the assumption that $+$ is chosen with probability
$p$ and $-$ with $1-p$. Other efforts to understand such sequences
can be found in \cite{makover2006elementary,krapivsky2001random,rittaud2007average}.

\subsection{Our Contribution}

To the best of the authors' knowledge, there has been no such study
of random Fibonacci sequences that take the form
\begin{equation}
\mathcal{I}_{n}=\eta_{n-1,n}\mathcal{I}_{n-1}+\eta_{n-2,n}\mathcal{I}_{n-2},\label{eq:system-1}
\end{equation}
where $\left\{ \eta_{i,j}\right\} $ is a set of strictly positive
i.i.d. random variables with continuous distribution where $\mathbb{E}\log\eta_{i,j}<\infty$, 
\[
\mathcal{I}_{1}:=\eta_{0,1}\mathcal{I}_{0},
\]
and $\mathcal{I}_{0}$ belongs to some bounded domain in $\mathbb{R}^{+}$. We call this type of sequence a continuous random Fibonacci sequence.
In this manuscript, we motivate the study of such sequences by demonstrating that the
rate at which the end-to-end capacity of a \emph{cooperative} AF
relay network will decay can be determined by the exponential growth rate of these sequences.

 {\subsection{Notation and Definitions\label{subsec:Notation}}}

Similar to notation used in \cite{simmons2015capacity}, for a strictly positive random variable $f(n)$ depending on~$n$, and some $h(n)=|o(n)|$,
\begin{eqnarray}
\!f(n)\!\!\!\!\!&=&\!\!\!\!\!O_{\mathbb P}\left(g(n)\right)\;\;\Rightarrow  \; \lim_{n\to\infty}\!\! \mathbb P\! \left[    f(n)   \leq\!    g(n) e^{h(n)}   \right] = 1\nonumber .\nonumber\\
\!f(n)\!\!\!\!\!&=&\!\!\!\!\!\Omega_{\mathbb P}\left(g(n)\right) \;\;\Rightarrow   \;\lim_{n\to\infty}\!\! \mathbb P\! \left[    f(n)   \geq \!    g(n) e^{-h(n)}   \right] = 1 \nonumber\\
\!f(n)\!\!\!\!\! &=&\!\!\!\!\! \Theta_{\mathbb P}\left( g(n)\right)\;\mathrm{if}\;f(n)=O_{\mathbb P}\left(g(n)\right)\;\mathrm{and}\;f(n)=\Omega_{\mathbb P}\left(g(n)\right).\nonumber
\end{eqnarray}

\section{Cooperative Network System Model\label{coopsystemMod}}

\begin{figure}
\centering{}\includegraphics[scale=0.55]{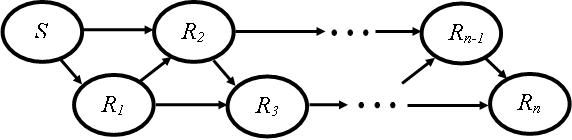}\caption{Depiction of a cooperative relay network.\label{fig:coop} }
\end{figure}

Consider a cooperative AF relay network, as depicted
in Fig. \ref{fig:coop}. The source node transmits an information symbol
with some magnitude $\mathcal{I}_{0}$ to the first and second relay.
The second relay waits for the first relay to amplify and transmit its received
signal from the source. Once this has occurred, the second and third
relays receive the transmission from the first node. After phase alignment,
the second relay sums the received signals from the source and first
relays, amplifies the sum, and transmits it to the third and fourth
relays. In general, the $i$th relay performs phase alignment on the
signals received from the $\left(i-1\right)$th and $\left(i-2\right)$th
nodes, amplifies it, and transmits it to the $\left(i+1\right)$th
and $\left(i+2\right)$th nodes. Consequently, the magnitude of the
transmitted information signal, $\mathcal{I}_{n}$, at the $n$th
node can be obtained from the random Fibonacci sequence given in \eqref{eq:system-1}, where $\eta_{i,j}=\left|h_{i,j}g_{j}\right|$, $h_{i,j}$ represents
the channel coefficient for the $\left(i,j\right)$th hop with $\mathbb{E}\left[ \vert h_{i,j}\vert^2 \right]=  \mu_{i,j}$, and $g_{j}$
represents the amplification factor applied at node $j$.

The information symbol transmitted at the $n$th node can be converted into the  following matrix expression
\begin{equation}
\left[\begin{array}{c}
\mathcal{I}_{n-1}\\
\mathcal{I}_{n}
\end{array}\right]=\prod_{i=2}^{n}\left[\begin{array}{cc}
0 & 1\\
\eta_{i-2,i} & \eta_{i-1,i}
\end{array}\right]\left[\begin{array}{c}
\mathcal{I}_{0}\\
\eta_{0,1}\mathcal{I}_{0}
\end{array}\right],\label{eq:Coop_Info_term-1}
\end{equation}
which is required so that we can leverage results from random dynamical systems (RDS) theory, \cite{arnold1998random}.
We further assume that a noise term $v_{j}\sim\mathcal{CN}\left(0,n_{0}\right)$
is received at the $j$th relay during the reception from the previous
two nodes. After averaging over the noise, the transmitted
noise power at the $n$th node, $\mathcal{N}_{n}^{2}$, can be obtained from the matrix system
\begin{equation}
\left[\begin{array}{c}
\mathcal{N}_{n-1}^{2}\\
\mathcal{N}_{n}^{2}\\
1
\end{array}\right]=\prod_{i=2}^{n}\left[\begin{array}{ccc}
0 & 1 & n_{0}\\
\eta_{i-2,i}^{2} & \eta_{i-1,i}^{2} & n_{0}\\
0 & 0 & 1
\end{array}\right]\left[\begin{array}{c}
0\\
0\\
1
\end{array}\right],\label{eq:TTSfib-1}
\end{equation}
so that the SNR at the $n$th node is given by 
\begin{equation}
\gamma_{n}:=\frac{\mathcal{I}_{n}^{2}}{\mathcal{N}_{n}^{2}},\label{eq:fibgamma}
\end{equation}
and the transmit power is given by 
\begin{equation}
X_{n}^2 = {\mathcal{I}_{n}^{2}} + {\mathcal{N}_{n}^{2}}.\label{eq:CoopTransPwr}
\end{equation}
With (\ref{eq:fibgamma}), we can write the capacity of the network 
as \cite{cover2012elements}
\begin{equation}
c_{n}=\log\left(1+\gamma_{n}\right)\;\mathrm{nats/channel\; use}.\label{eq:fibcapacity}
\end{equation}

\section{Capacity and Power Scaling Laws for Cooperative Relay Networks}

In this section, we establish capacity and power scaling laws for the network described in section \ref{coopsystemMod}. We begin with capacity in the following theorem.
\begin{theorem}
\textup{\label{Thm:Fibnacc}}
Let $\lambda$ be the upper Lyapunov exponent of the RDS described in \eqref{eq:Coop_Info_term-1}. Then the capacity of the network is given by
\begin{equation}
c_n = \Theta \left(     \exp\left(       n\min\left\{0 , 2 \lambda  \right\}  \right)    \right).
\end{equation}
\end{theorem}
\begin{IEEEproof}
The theorem follows from \cite[Theorem 1]{simmons2015capacity}, which shows that 
$
\lambda_\mathcal{N} := \lim_{n\to\infty}\frac{1}{n}\log  \mathcal{N}_n^2 ,
$ is given by
\begin{equation}
\lambda_\mathcal{N} \overset{a.s.}{=} \max\left\{   0   ,   2\lambda  \right\}.\label{eq:noiseLya}
\end{equation}
From \eqref{eq:fibgamma} and \eqref{eq:noiseLya}, we obtain
\begin{align}
\lim_{n\to\infty}\frac{1}{n}\log \gamma_n  & = \lim_{n\to\infty}\frac{1}{n}\log  \mathcal{I}_n^2   -   \lim_{n\to\infty}\frac{1}{n}\log   \mathcal{N}_n^2\\
& = 2\lambda  -  \max\left\{  0 , 2\lambda  \right\}= \min\left\{  0   , 2\lambda \right\}   .
\end{align}
The existence of these limits follows from the Furstenberg-Kesten theorem. By applying the same technique that was used in \cite[Theorem 2.B]{simmons2015capacity}, it is easy to show that
\begin{align}
\lim_{n\to\infty}\frac{1}{n}\log c_n  & = \min\left\{  0   , 2\lambda \right\}   .\label{eq:capacity_proof}
\end{align}
The stated result follows from \cite[Fact 1.4]{simmons2015capacity}.
\end{IEEEproof}

In the next lemma, we will determine the power scaling properties of the network. With this, we will be able to establish the relationship between the decay of the capacity across the network and the growth in its transmit power. In particular, we will conclude that it is possible to construct cooperative AF networks that avoid both exponential capacity decay and exponential transmit power growth.
\begin{lemma}
\textup{\label{Thm:Fibnacc2}}
Let $\lambda$ be the upper Lyapunov exponent of the random dynamical system described in \eqref{eq:Coop_Info_term-1}. Then the transmit power at the $n$th node is given by
\begin{equation}
X_n^2 = \Theta \left(     \exp\left(       n\max\left\{0 , 2 \lambda  \right\}  \right)    \right).
\end{equation}
\end{lemma}
\begin{IEEEproof}
From \eqref{eq:CoopTransPwr}, it is clear that
\begin{align}
 \frac{1}{n}\log   \mathcal{N}_n^2   \leq \frac{1}{n}\log X_n^2  \leq \frac{1}{n}\log  2\max\left\{\mathcal{I}_n^2,\mathcal{N}_n^2\right\}.   \label{eq:proofLyaXn}
\end{align}
Taking the limit of \eqref{eq:proofLyaXn} yields
\begin{align}
\lim_{n\to\infty} \frac{1}{n}\log X_n^2  =  \max\left\{   0   ,   2\lambda  \right\} \label{eq:proofLyaXn2}.
\end{align}
The stated result then  follows from \cite[Fact 1.4]{simmons2015capacity}.
\end{IEEEproof}

Theorem \ref{Thm:Fibnacc} and Lemma \ref{Thm:Fibnacc2} demonstrate that it is possible to construct a cooperative AF network that simultaneously avoids exponential capacity decay and exponential transmit power growth. To achieve this, the network must be constructed such that its Lyapunov exponent is zero; i.e., $\lambda = 0$.

\section{Conclusion}

We have motivated the study of continuous random Fibonacci sequences. In particular, we have shown that the Lyapunov exponent of these sequences can be used to establish capacity and power scaling laws for cooperative AF relay networks. Furthermore, with the laws that we have established, we are able to see that cooperative AF networks can be constructed that avoid both exponential capacity decay and exponential transmit power growth across the network. This is achieved by ensuring that the Lyapunov exponent of the sequence is zero. Future work may focus on calculating the Lyapunov exponent.

\bibliographystyle{plain}
\bibliography{fibBib}

\end{document}